\documentclass[final,5p,times,twocolumn]{elsarticle}
\usepackage{amsmath}
\usepackage{latexsym}
\usepackage[bf]{caption}
\usepackage{indentfirst}
\usepackage{graphicx}
\usepackage{epsf}
\usepackage{psfrag}

\begin{document}

\begin{frontmatter}

\title{Final state interactions in $B^{\pm} \rightarrow K^+ K^- K^{\pm}$ decays}

\author[b]{A. Furman}
\author[a]{R. Kami\'nski}
\author[a]{L. Le\'{s}niak\corref{LL}}
\ead{Leonard.Lesniak@ifj.edu.pl}
\cortext[LL]{Corresponding author}
\author[a]{P. \.Zenczykowski}
\address[b]{ul. Bronowicka 85/26, 30-091 Krak\'ow, Poland}
\address[a]{Henryk Niewodnicza\'nski Institute of Nuclear Physics, Polish Academy of Sciences, \\~~~~PL 31-342 Krak\'ow, Poland}

\begin{abstract}
Charged  $B$ decays to three charged kaons are analysed  
in the framework of the QCD 
factorization approach. The strong final state $K^+K^-$
interactions  are described 
using the kaon scalar and vector form factors. The
 scalar non-strange and strange form factors at low 
$K^+K^-$ effective masses are constrained by chiral 
perturbation theory and satisfy the two-body unitarity
conditions. The latter stem from the properties of the meson-meson 
amplitudes which describe all possible $S$-wave transitions between
three coupled channels consisting of two kaons, two pions
and four pions. The vector form factors are fitted 
to the data on the electromagnetic kaon interactions.
   The model results are compared with the Belle and 
BaBar data. Away from $\phi(1020)$ resonance, in the $S$-wave 
dominated $K^+K^-$ mass spectra, a possibility for a large $CP$ asymmetry 
is identified.
\end{abstract}

\end{frontmatter}

\section{\bf Introduction}

Recently, charmless three-body decays of $B$ mesons have been 
intensively studied both experimentally and theoretically.
On the experimental side,
Dalitz plot analyses of the charged $B$ decays were performed
by Belle \cite{Belle} and BaBar \cite{BaBar} collaborations. Likewise, several 
theoretical studies involving the $B^{\pm} \rightarrow K^+ K^- K^{\pm}$ decays
have been published~\cite{Fajfer},~\cite{Furman} and ~\cite{Cheng}.

Since charged kaons interact strongly, their long distance interactions
in the final states have to be well understood if one aims at extracting weak decay amplitudes 
from the $B$ to $KKK$ decays. 
In this Letter we go beyond an isobar model parameterization
of the $B$ decay amplitudes and introduce additional theoretical constraints on the
$S$-wave two-body $K^+K^-$ interaction amplitudes, which follow, in particular, from unitarity.
In order to satisfy unitarity in two-body interactions we construct scalar strange 
and non-strange form factors which enter into the $S$-wave parts of the decay amplitudes.
These amplitudes are calculated in the framework of the QCD factorization 
approach. In the  construction of form factors we use experimental information on
the $K^+K^-$ interactions coming from experiments other than $B$ decays, for example from
$K^+K^-$ production processes in hadronic collisions or from $e^+e^-$ reactions.
We apply also some low-energy constraints coming from the chiral perturbation theory.
Preliminary results of our analysis concerning the  
$B^{\pm} \rightarrow K^+ K^- K^{\pm}$ reactions can be found in Ref.~\cite{ICHEP}.

In Section~2 we formulate the theoretical model of the $B^+$ and $B^-$ decay 
amplitudes. Presentation of results and their comparison with the experimental data are given 
in Sec.~3. Our conclusions are presented in Sec.~4.

\section{\bf $B^{\pm} \rightarrow K^+ K^- K^{\pm}$ decay amplitudes}

Inspection of the Dalitz plots of the Belle \cite{Belle} and BaBar
 \cite{BaBar} experiments reveals an accumulation
of events for the $K^+K^-$ effective masses below 1.8 GeV.
Indeed, several mesonic resonances which can decay into the $K^+K^-$
pairs  exist in this range \cite{PDG}. Among them there are
scalar and vector resonances which are formed via the $S$- and $P$-wave
final state interactions. In the first approximation one can neglect 
their interaction with the third kaon. This justifies 
using the QCD quasi-two-body factorization
approach for the limited range of the effective $K^+ K^-$ masses (see, for example Ref. \cite{Beneke}).
The $B^- \rightarrow K^+ K^- K^-$ amplitude is then expressed in terms of the following matrix element of the weak 
effective Hamiltonian $H$:
\begin{equation}
\label{A}
\langle K^-(p_1)K^+(p_2)K^-(p_3)|H|B^-\rangle= A^-_S + A^-_P,
\end{equation}
where the $S$-wave part is
\begin{equation}
\label{AS}
\begin{split}
A^-_S= &\dfrac{G_F}{\sqrt{2}}\Big\{-\sqrt{\dfrac{1}{2}}\chi
 f_{K}(M^2_B-s_{23})F^{B\to(K^+K^-)_S}_0(m^2_{K})y
\Gamma^{n^*}_2(s_{23})
\\&
+\dfrac{2B_0}{m_b-m_s} (M^2_B-m^2_{K})F^{B K}_0(s_{23})v
\Gamma^{s^*}_2(s_{23})\Big\},
\end{split}
\end{equation}
the $P$-wave part is
\begin{equation}
\label{AP}
\begin{split}
 A^-_P=&\dfrac{G_F}{\sqrt{2}}\Big\{\dfrac{f_K}{f_{\rho}}A^{B\rho}_0(m^2_{K})y F^{K^+K^-}_u(s_{23})
-F^{BK}_1(s_{23})\Big[w_uF^{K^+K^-}_u(s_{23})
\\&
+w_dF^{K^+K^-}_d(s_{23})+w_sF^{K^+K^-}_s(s_{23})
\Big]\Big\}4\overrightarrow{p}_1 \cdot \overrightarrow{p}_2 
\end{split}\end{equation}
and the interacting kaons are taken to be kaons 2 and 3. Furthermore,
$s_{23}$ is the square of the $K^+(p_2)K^-(p_3)$ effective mass 
$m_{23} \equiv m_{K^+K^-}$, while
$\overrightarrow{p}_1$ and $\overrightarrow{p}_2$ are the kaon 1 and kaon 2 momenta in the center
of mass system of the kaons 2 and 3. The scalar product of the kaon momenta can be written in terms of the
helicity angle $\Theta_H$:
\begin{equation}
\label{p1p2}
\overrightarrow{p}_1 \cdot \overrightarrow{p}_2=-|\overrightarrow{p}_1||\overrightarrow{p}_2|
\cos \Theta_H.
\end{equation}
In these equations $G_F$ is the Fermi coupling constant, $f_K=0.1555$ GeV and 
$f_{\rho}=0.220$ GeV are the kaon and the $\rho$ meson decay constants,
$M_B$, $m_K$, $m_b=4.9$ GeV, $m_s=0.1$ GeV, $m_u= 0.004$ GeV and 
$m_d = 0.004$ GeV are the masses of the $B$ meson, kaon, $b$-quark, strange
 quark, down- and up-quarks, respectively.
  
The functions 
$\Gamma^{n}_2$ and $\Gamma^{s}_2$, present in the $S$-wave amplitude in Eq.~(\ref{AS}), are the kaon 
non-strange and strange scalar form factors. The vector form factors
 $F^{K^+K^-}_q$ (for $q=u,d$ and $s$), introduced in Eq.~(\ref{AP}), are defined through matrix 
elements
 \begin{equation}
\label{Fq}
<K^+(p_2)K^-(p_3)|\bar q \gamma_{\mu} q |0>=(p_2-p_3)_{\mu}F^{K^+K^-}_q(s_{23}),
\end{equation}
where $|0>$ is the vacuum state.
The $K^+K^-$ pair in the $S$-wave is then denoted by $R_S \equiv (K^+K^-)_S$. Similarly  
$R_P \equiv (K^+K^-)_P$ stands for the $P$-state. Furthermore 
$F^{B\to(K^+K^-)_S}_0$ in Eq.~(\ref{AS}) is the form factor of the transition from the $B$ meson 
to the $K^+K^-$ pair in the $S$-state, 
$\chi$ is the constant related to the decay of the $(K^+K^-)_S$ state into two kaons, and
$B_0= m_{\pi}^2/(m_u+m_d)$, where $m_{\pi}$ is the pion mass.
We take $F^{B\to(K^+K^-)_S}_0(m_K^2)=0.13$ \cite{Bruno} and we fit $\chi$ to the data.
Functions $F^{B K}_0(s_{23})$ and  $F^{B K}_1(s_{23})$ are the $B\to K$ scalar and vector transition form
 factors and $A^{B\rho}_0(m_K^2)=0.37$ \cite{Beneke} is the $B\to\rho$ transition form factor. 
In our approximation, the ratio $A^{B\rho}_0/f_{\rho}$ 
represents a general factor related to the transition from $B^-$ to any $(K^+K^-)_P$ state 
and then its decay into the final $K^+K^-$ pair. For the case of the $\rho$ meson 
this coupling to the pair of kaons is effectively realized only above the $K^+K^-$ threshold.

The weak decay amplitudes depend on QCD factorization coefficients $a_j^p$ and on the products 
$\Lambda_u=V_{ub}V_{us}^*$, $\Lambda_c=V_{cb}V_{cs}^*$, where $V_{ij}$ 
are the CKM quark-mixing matrix elements. In order to describe $B$ decay into mesons $M_1$ and $M_2$ 
we follow Ref.~\cite{Beneke} and calculate
the coefficients $a_j^p(M_1~M_2)$ at the next-to-leading order in the strong coupling constant
at the renomalization scale equal to $m_b/2$. Here the $M_1$ meson has a common spectator quark with the
decaying $B$ meson. In the case of the $B^- \rightarrow K^+ K^- K^-$ decays, $M_1$ or $M_2$ can be 
either kaon $K^-$, or systems $R_S$, $R_P$. We take into account one-loop vertex
and penguin corrections to $a_j^p(M_1M_2)$  but neglect those due to  hard scattering or the 
annihilation since they are expected to be generally suppressed. In the QCD factorization approach they 
receive logarithmically divergent contributions due to soft gluon interaction which are
``unavoidably model dependent'' (see Ref.~\cite{Beneke}). We treat such soft interactions by 
introducing the form factors constrained by data on meson-meson interactions, taken from analyses of
reactions other than the $B$ decays. Under these conditions
we have $a_j^p(R_SM_2)=a_j^p(R_PM_2)$, with their common value denoted below by 
$a_j^p(R_{S,P}M_2)\equiv a_{jy}^p$. We also use the abbreviations: $a_{jw}^p\equiv a_j^p(K^-R_P)$
and $a_{jv}^p\equiv a_j^p(K^-R_S)$.
The values of coefficients $a_j^p(M_1M_2)$ are given in Table~\ref{Tab1}.

In terms of the quantities introduced above one defines:
\begin{equation}
\label{y}
\begin{split}
 y=\Lambda_u\Big[a_{1y}+a_{4y}^u+a_{10y}^u-(a_{6y}^u
+a_{8y}^u)r_{\chi}^K\Big]+
\\
\Lambda_c\Big[a_{4y}^c+
a_{10y}^c-(a_{6y}^c+a_{8y}^c)r_{\chi}^K\Big],
\end{split}
\end{equation}
where
\begin{equation}
\label{rchi}
 r_{\chi}^K=\dfrac{2m_K^2}{(m_b+m_u)(m_u+m_s)},
\end{equation}

\begin{equation}
\label{wu}
 w_u=\Lambda_u(a_{2w}+a_{3w}+a_{5w}+a_{7w}+
a_{9w})+\Lambda_c(a_{3w}+a_{5w}+a_{7w}+a_{9w}),
\end{equation}

\begin{equation}
\label{wd}
 w_d=\Lambda_u\Big[a_{3w}+a_{5w}-\dfrac{1}{2}(a_{7w}+
a_{9w})\Big]+\Lambda_c\Big[a_{3w}
+a_{5w}-\dfrac{1}{2}(a_{7w}+a_{9w})\Big],
\end{equation}

\begin{equation}
\label{ws}
\begin{split}
 w_s=\Lambda_u\Big[a_{3w}+a_{4w}^u+a_{5w}-\dfrac{1}{2}(a_{7w}
+a_{9w}+a_{10w}^u)\Big]+
\\
\Lambda_c\Big[a_{3w}+a_{4w}^c+
a_{5w}-\dfrac{1}{2}(a_{7w}+a_{9w}+a_{10w}^c)\Big],
\end{split}
\end{equation}
and
\begin{equation}
 \label{v}
v=\Lambda_u(-a_{6v}^u+\dfrac{1}{2}a_{8v}^u)+\Lambda_c(-a_{6v}^c+
\dfrac{1}{2}a_{8v}^c).
\end{equation}

One can notice that in the expressions for the decay amplitudes there are no transitions to the 
$K^+K^-$ states
of spin 2 or higher. This results from the application of the factorization approach in which
matrix elements to spin states higher than one vanish. The contribution of $f_2(1270)$ with its
 rather small branching fraction to $K \bar K$ (4.6 \%) is thus not included in this study.

Since two identical charged kaons appear in the final state of the $B^- \rightarrow K^+ K^- K^-$ decay,
the amplitude of  Eq.~(\ref{A}) has to be symmetrized
\begin{equation}
 \label{Asym}
\begin{split}
A_{sym}^-=&\dfrac{1}{\sqrt{2}}\Big[\langle K^-(p_1)K^+(p_2)K^-(p_3)|H|B^-\rangle+\\&
\langle K^-(p_3)K^+(p_2)K^-(p_1)|H|B^-\rangle \Big].
\end{split}
\end{equation}
The symmetrized amplitude for the  $B^+ \rightarrow K^+ K^- K^+$ reaction reads
\begin{equation}
 \label{Aplus}
A_{sym}^+=A_{sym}^-(\Lambda_u \rightarrow \Lambda_u^*,\Lambda_c \rightarrow \Lambda_c^*, 
B^- \rightarrow B^+).
\end{equation}

The final state kaon-kaon $S$-wave interactions are dynamically coupled with systems consisting of
 two and four pions. Thus a system of three coupled 
channels: $\pi\pi, \bar KK$ and $4\pi$ (effective $(2\pi)(2\pi)$ or $\sigma\sigma$, $\rho\rho$ etc.),
 labelled by $j=1,2,3,$ is 
considered in the construction of scalar form factors $\Gamma^{n}_2$ and $\Gamma^{s}_2$.
Here we use an approach initiated in \cite{Furman} and recently developed in \cite{3pi} for 
 the $B^{\pm} \rightarrow \pi^+ \pi^- \pi^{\pm}$ decays. A set of the 3x3 transition 
amplitudes $T$, describing all possible transitions between the three channels, is taken from a unitary
 model of Ref. \cite{KLL} (solution $A$). We introduce two kinds of production functions
 $R_j^{n,s}$, labeled by $n$ (non-strange) or by $s$ (strange): 
\begin{equation}
\label{R}
R_j^{n,s}(E)=\frac{\alpha_j^{n,s}+\tau_j^{n,s}E+\omega_j^{n,s}E^2}{1+cE^4},~~~j=1,2,3,
\end{equation}
where $\alpha_j^{n,s},\tau_j^{n,s},\omega_j^{n,s}$ and $c$ are constant parameters, while $E$ represents 
the total energy and is related to the center of mass momenta $k_j=\sqrt{E^2-m_j^2}$, with 
 $m_1=m_\pi$, $m_2=m_K$, $m_3=700$ MeV, and $s\equiv E^2\equiv m_{K^+K^-}^2$.
 The three scalar form factors, written in the compact row matrix form $\Gamma^{n,s*}$, are given by
\begin{equation}
\label{macG}
\Gamma^{n,s*}=R^{n,s}+TGR^{n,s},
\end{equation} 
where $R^{n,s}$ are rows of the production functions and $G$ is the matrix of the Green's functions
multiplied by the convergence factors $F_j(p)=(k_j^2+\kappa^2)/(p^2+\kappa^2)$. These factors,
which reduce to unity on shell ($p=k_j$), make finite the relevant integrals over the intermediate
momenta $p$. The parameter $\kappa$ will be fitted to the data of the 
BaBar \cite{BaBar} and Belle \cite{Belle} Collaborations.

 For both the non-strange and strange form
 factors we also constrain
their low energy behaviour using the chiral perturbation model of Refs.~\cite{MO,Lahde}.
 At low $s$ values one writes the following expansion:
\begin{equation}
\label{lowgamma}
\Gamma_j^{n,s}(s) \cong d_j^{n,s}+f_j^{n,s}s, \quad j=1,2,3,  
\end{equation}
with real coefficients $d_j^{n,s}$ and $f_j^{n,s}$.
Explicit formulae for the set of non-strange form factors, in particular for the $\Gamma^n_2$ 
presented in Eq.~(\ref{AS}), are given in Eqs. (24-35) of Ref.~\cite{3pi}.
For the strange form factors we have
\begin{equation}
\label{d1s}
 d_1^{s}=\frac{\sqrt{3}}{2} \left [\frac{16m_\pi^2}{f^2}\left( 2L_6^r-L_4^r\right ) 
          -\frac{m_\pi^2}{72\pi^2 f^2}\left(1+\log{\frac{m_\eta^2}{\mu^2}}\right )\right],
 \end{equation}
         
\begin{equation}
\label{f1s}
 f_1^{s}= \frac{\sqrt{3}}{2} \left [ \frac{8L_4^r}{f^2}-
\frac{1}{32\pi^2 f^2}\left(1+\log{\frac{m_K^2}{\mu^2}}\right)
+\frac{m_\pi^2}{432\pi^2m_\eta^2f^2}\right],
\end{equation}
and
\begin{equation}
\label{d2s}
\begin{split}
 &d_2^{s}= 1+ \frac{8(2L_6^r-L_4^r)}{f^2}\left(m_{\pi}^2+4m_K^2\right)
       - \frac{16L_5^r}{f^2}m_K^2
        \\&+ \frac{32L_8^r}{f^2}m_K^2
+\frac{m_\eta^2}{48\pi^2f^2}\log{\frac{m_\eta^2}{\mu^2}}
        +\frac{m_K^2}{36\pi^2f^2}\left(1+\log{\frac{m_\eta^2}{\mu^2}}\right),
\end{split}
\end{equation}
\begin{equation}
\label{f2s}
\begin{split}
 f_2^{s}=\frac{8L_4^r}{f^2}&+\frac{4L_5^r}{f^2}-\frac{m_K^2}{216\pi^2f^2m_\eta^2}
           -\frac{1}{32\pi^2f^2}\left(1+\log{\frac{m_\eta^2}{\mu^2}}\right)
\\&
-\frac{3}{64\pi^2f^2}\left(1+\log{\frac{m_K^2}{\mu^2}}\right).
\end{split}
\end{equation}
In these equations $m_\eta$ is the $\eta$ meson mass, $\mu$ is the scale of the dimensional 
regularization and $f=f_{\pi}/\sqrt{2}$. Using $f=92.4$~MeV and the chiral perturbation theory
 constants
 $L_k^r$, $k=4,5,6,8$, given in Table X of Ref.~\cite{Allton}, we obtain the 
non-strange-sector parameters: $d_1^n=1.1957$, $f_1^n=3.1329$~GeV$^{-2}$,
 $d_2^n=0.7193$ and $f_2^n=1.6719$~GeV$^{-2}$ and their strange-sector counterparts:
$d_1^s=-0.0016$, $f_1^s=0.2393$~GeV$^{-2}$, $d_2^s=1.0410$ and $f_2^s=0.6235$~GeV$^{-2}$.
For the form factors related to the third channel at low energies
we make the simplest assumptions 
$d_3^{n}=d_3^{s}=f_3^{n}=f_3^{s}=0$, as in Ref.~\cite{Bachir}.

 The coefficients $\alpha_j^{n,s},\tau_j^{n,s}$ and $\omega_j^{n,s}$
are constrained by the values of the form factors at low energies.
 They are calculated using the low energy expansion of Eq.~(\ref{macG})
and are listed in Table~\ref{Tab.a.tal.omega}. The parameter $c$, 
which controls the high energy behaviour of $R$, is fixed while fitting the data.

Our scalar form factors satisfy the following unitarity conditions:
\begin{equation}
\label{unit}
Im~ \Gamma^* = T^{\dagger}D~\Gamma^*,
\end{equation}
where $D$ is the diagonal matrix of the kinematical coefficients which are proportional to the channel
 momenta $k_j$ in the center of mass frame:
\begin{equation}
\label{Dij}
D_{ij}=-\frac{k_j\sqrt{s}}{8\pi}\delta_{ij}~\theta~(\!\sqrt{s}-2m_j),~~~~~i,j=1,2,3.
\end{equation}

\begin{table*}
\caption{
\label{Tab1} 
Leading order (LO) and next-to-leading order (NLO) coefficients  $a_{iy}^{p}$,
 $a_{iv}^{p}$ and  $a_{iw}^{p}$ entering into Eqs.~(\ref{y}-\ref{v}).
The NLO coefficients are the sum of the LO coefficients plus next-to-leading order vertex
 and penguin corrections.
The superscript $p$  is omitted for $ i=1$, $2$, $3$, $5$, $7$ and $9$,
 the penguin corrections being zero for these cases.
}
\begin{tabular}{ccccccc}

\hline
            &\multicolumn{2}{c}{$a_{iy}^{p}$~~~~~~~~~} &\multicolumn{2}{c}{$a_{iv}^{p}$~~~~~~~~~ }&\multicolumn{2}{c}{$a_{iw}^{p}$~~~~~~~~~ } \\
            &       LO     &    NLO                         &      LO    &  NLO                   &      LO    &  NLO                 \\
\hline
$a_1$       & $1.039$      & $1.066  + i 0.039$                &            &                           &            &     \\                        
$a_2$       &              &                                   &            &                           &  $0.084$   &  $-0.041 -i 0.114$    \\      
$a_3$       &              &                                   &            &                           &  $0.004$   &  $0.010 -i 0.005$    \\      
$a_4^{u}$   &$    -0.044 $ & $-0.029  - i 0.02$                &            &                           &  $-0.044$  & $-0.032  - i 0.019$  \\       
$a_4^c$     & $ -0.044 $   & $ -0.035  - i 0.004$              &            &                           &  $-0.044$  &  $ -0.038  - i 0.006$\\       
$a_5$       &              &                                   &            &                           &  $-0.012$  &  $-0.010 -i 0.007$    \\      
$a_6^{u}$   & $-0.062 $    & $ -0.057  - i 0.017  $            & $-0.062$   & $-0.075-i 0.017$          &            &    \\                         
$a_6^c$     &$  -0.062 $   & $ -0.062  - i 0.004$              & $-0.062$   & $-0.079-i 0.004$          &            &                        \\     
$a_7$       &              &                                   &            &                           &  $0.0001$  &  $0.0 +i 0.0001$   \\         
$a_8^{u}$   &$0.0007$      & $ 0.0008 + i 0.0$                 & $0.0007$   & $0.0007+i 0.0$            &            &  \\                           
$a_8^c$     &$0.0007$      & $0.0008 + i 0.0 $                 & $0.0007$   & $0.0006+i 0.0$            &            &                        \\     
$a_9$       &              &                                   &            &                           &  $-0.0094$ &  $-0.0097 -i 0.0003$    \\    
$a_{10}^{u}$& $-0.0009 $   & $0.0005 + i 0.0013$               &            &                           &  $-0.0009$ &  $0.0006 + i 0.001$   \\      
$a_{10}^c$  &  $-0.0009$   & $0.0005 + i 0.0013$               &            &                           &  $-0.0009$ &  $0.0006 + i 0.001$   \\
\hline
\end{tabular}
\end{table*}

Presence of the resonances in the $K^+K^-$ effective mass distributions 
(see Refs.~\cite{BaBar,Belle}) is a direct manifestation of the $K^+K^-$ final state 
interactions. The most prominent resonance in the $P$-wave is $\phi(1020)$.
In 2005 Bruch, Khodjamirian and K\"{u}hn~\cite{Bruch} described the 
electromagnetic 
form factors for charged and neutral kaons in terms of additive contributions from eight vector 
mesons: $\rho \equiv \rho(770)$, $\rho^{'} \equiv \rho(1450)$, $\rho^{''} \equiv \rho(1700)$, 
$\omega \equiv \omega(782)$, 
$\omega ^{'} \equiv \omega(1420)$, 
$\omega^{''} \equiv \omega(1650)$, $\phi \equiv \phi(1020)$ and $\phi{'} \equiv \phi(1680)$. 
Using quark model assumptions 
and isospin symmetry as in Ref.~\cite{Bruch} one can deduce the following expressions for the three $P$-wave form factors
$F^{K^+K^-}_q$ defined in Eq.~(\ref{Fq}):
\begin{equation}
\label{Fu}
\begin{split}
F^{K^+K^-}_u&=\frac{1}{2}(c_{\rho}BW_{\rho}+c_{\rho^{'}}BW_{\rho^{'}}+c_{\rho^{''}}BW_{\rho^{''}}
\\&
+c_{\omega}BW_{\omega}+c_{\omega^{'}}BW_{\omega^{'}}+c_{\omega^{''}}BW_{\omega^{''}}),
\end{split}
\end{equation}
\begin{equation}
\label{Fd}
\begin{split}
F^{K^+K^-}_d&=\frac{1}{2}(-c_{\rho}BW_{\rho}-c_{\rho^{'}}BW_{\rho^{'}}-c_{\rho^{''}}BW_{\rho^{''}}
\\&
+c_{\omega}BW_{\omega}+c_{\omega^{'}}BW_{\omega^{'}}+c_{\omega^{''}}BW_{\omega^{''}}),
\end{split}
\end{equation}
\begin{equation}
\label{Fs}
F^{K^+K^-}_s=-c_{\phi}BW_{\phi}-c_{\phi{'}}BW_{\phi{'}}.
\end{equation}
In the above equations $BW_i$, i=1,...,8,  are the energy-dependent Breit-Wigner functions, defined
for each resonance of mass $m_i$ and width $\Gamma_i$ as
\begin{equation}
\label{BW}
BW_i(s)=\frac{m_i^2}{m_i^2- s-i\sqrt{s}~\Gamma_i(s)},
\end{equation}
and $c_i$ are the constants given in Table 2 of Ref.~\cite{Bruch} for the constrained fit.

   The $B$ to $K$ transition form factors have been parametrized according to Ref.~\cite{Ball}:
\begin{equation}
\label{F0BK}
F_0^{BK}(s)=\frac{r_0}{1-\frac{s}{s_0}}, 
\end{equation}
where $r_0=0.33$, $s_0=37.46$ GeV$^2$, and
\begin{equation}
\label{F1BK}
F_1^{BK}(s)=\frac{r_1}{1-\frac{s}{m_1^2}}+\frac{r_2}{(1-\frac{s}{m_1^2})^2}, 
\end{equation}
where $r_1=0.162$, $r_2=0.173$ and $m_1=5.41 $ GeV.
\begin{table*}
\caption{\label{Tab.a.tal.omega}Parameters of production functions $R_i^n(E)$ and $R_i^s(E)$ 
defined in Eq.~(\ref{R}) for  $\kappa$ =3.506 GeV}
\begin{tabular}{ccccccc}
\hline 
     i & $\alpha_i^n$ & $\tau_i^n$ (GeV$^{-1}$) & $\omega_i^n$ (GeV$^{-2}$) & $\alpha_i^s$ & $\tau_i^s$ (GeV$^{-1}$) & $\omega_i^s$ (GeV$^{-2}$)\\
\hline 
     1 &  $ 0.6731$   & $-0.2511$               & $ 1.5301$                 &  $ 0.3743 $  & $-0.1090$               & $ 0.1008$\\
 
     2 &  $0.6116$    & $ 0.0428$               & $ 1.5232$                 &  $ 0.7075 $  & $-0.1029$               & $ 0.3256$\\ 
 
     3 &  $1.2055$    & $ 0.3589$               & $ 3.1556$                 &  $ 1.0028 $  & $+0.0979$               & $ 0.4653$ \\ 
\hline 
\end{tabular}
\end{table*}
\section{\bf Results}
 Partial wave analysis of the decay 
amplitudes helps in the investigation of the density distributions in the Dalitz diagrams.
In Eqs.~(\ref{AS},\ref{AP}) we have defined the $S$- and $P$- wave amplitudes to which 
the double differential $B^{-} \rightarrow K_1^- K_2^+ K_3^{-}$ branching fraction $Br$
is related through the symmetrized amplitude $A^-_{sym}$ of Eq.~(\ref{Asym}):
\begin{equation}
\label{d2Br}
\frac{d^2Br^-}{dm_{23}d \cos \Theta_H}= \frac{1}{\Gamma_B}\frac{m_{23}
\vert \overrightarrow{p_1} \vert \vert \overrightarrow{p_2} \vert }{8 (2\pi)^3 M_B^3} 
\left \vert A^-_{sym}(m_{23},\Theta_H)\right \vert^2.
\end{equation}
Here $\Gamma_B$ is the total width of the $B^-$ meson and the kaon momenta are:
\begin{equation}
\label{p1}
\vert \overrightarrow{p_1} \vert  = \frac{1}{2}\sqrt{m_{23}^2-4m_K^2},
\end{equation}
\begin{equation}
\label{p2}
\vert \overrightarrow{p_2} \vert = \frac{1}{2m_{23}}\sqrt{\left[M_B^2-\left(m_{23}+
m_K\right)^2\right]\left[M_B^2-\left(m_{23}-m_K\right)^2\right]}.
\end{equation}
The helicity angle $\Theta_H$ is kinematically related to the effective mass $m_{12}$ of the
$K^-_1 K^+_2$ system: 
\begin{equation}
\label{costetaH}
\cos \theta_H =\frac{1}{2\vert \overrightarrow{p_1} \vert  \vert \overrightarrow{p_2} \vert} 
\left[m_{12}^2-\frac{1}{2}\left(M_B^2-m_{23}^2+3m_K^2\right)\right].
\end{equation}
Due to the symmetry of the Dalitz plot density under the exchange of the kaons $K_1^-$ and $K_3^-$,
one can define the effective mass $m_{23}$ distribution integrated over the $m_{12}$ masses
larger than $m_{23}$:  
\begin{equation}
\label{dBrdm23}
\frac{dBr^-}{dm_{23}}=\int_{\cos \Theta_g}^{1}\frac{d^2Br^-}{dm_{23}d \cos \Theta_H}
 d\cos \Theta_H,
\end{equation}
where $\cos \Theta_g$ corresponds to the value of $\cos \Theta_H$ in Eq.~(\ref{costetaH})
 with $m_{12}=m_{23}$.
The helicity angle distribution $dBr^-/d \cos \Theta_H$ can be obtained from Eq.~(\ref{d2Br})
by integration over the specific range of the effective mass $m_{23}$. 

Our aim is to describe the data of the Belle~\cite{Belle} and BaBar~\cite{BaBar} Collaborations 
in one common fit. The data chosen by us include the total branching fraction for the decay 
$B^{\pm} \rightarrow [\phi(1020) K^{\pm}, \phi(1020) \rightarrow K^+K^-]$,
the averaged effective mass distributions $dBr^{\pm}/dm_{23}$ for  $m_{23}$ smaller than 1.8 GeV,
and the averaged helicity angle distribution $dBr^{\pm}/d \cos \Theta_H$ for $m_{23}<1.05$ GeV. 
The distributions of the $B^{\pm} \rightarrow K^+ K^- K^{\pm}$ events 
are obtained from the published data by subtraction of the background components. The total number
of data points for ten plots from both collaborations is equal to 175. The theoretical
distributions are normalized to the total number of experimental events corresponding to each 
data set. In our fit we used the averaged 
$B^{\pm} \rightarrow [\phi(1020) K^{\pm}, \phi(1020) \rightarrow K^+K^-]$ branching fraction equal
to $(4.06 \pm 0.34) \cdot 10^{-6}$ \cite{PDG}. There are four fitted parameters:
$\chi, \kappa, c$ and $N_P$. The first three parameters are related to the $S$-wave decay amplitudes
and the fourth one, $N_P$, is the common $P$-wave normalization constant by which the 
amplitudes $A_P^-$ and $A_P^+$ are multiplied.
\begin{figure}[t!]
\includegraphics[width=0.45\textwidth]{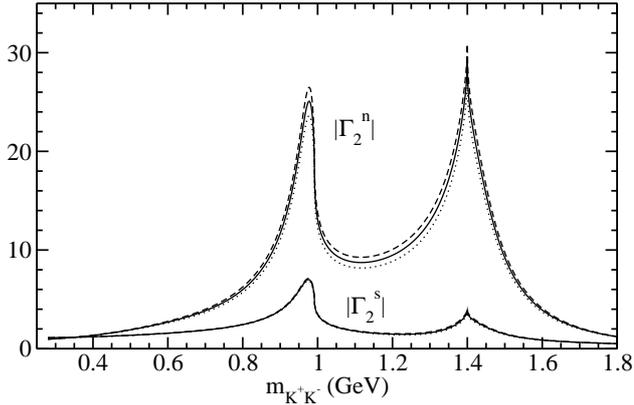}
\vspace*{6pt}

\caption{ Moduli of kaon scalar non-strange and strange form factors (solid lines) obtained in our fit.
The dashed and dotted lines represent the variation of their moduli
when parameter $\kappa$ varies within its error band.}
\end{figure}
   We have performed the fit to the 176 data points obtaining the total value of $\chi^2$ equal to 343
and the following parameters: 
$\chi=(6.44 \pm 0.44)$ GeV$^{-1}$, $\kappa=(3.51 \pm 0.20)$ GeV, $c=(0.084 \pm 0.010)$ GeV$^{-4}$
and $N_P=1.037 \pm 0.014$. For $N_P=1$ we obtain the averaged 
$B^{\pm} \rightarrow [\phi(1020) K^{\pm}, \phi(1020) \rightarrow K^+K^-]$ branching fraction
 equal to 3.73$\cdot 10^{-6}$ which is within one standard deviation from the experimental
 value of $(4.06 \pm 0.34) \cdot 10^{-6}$.
One sees that the absolute normalization of the $P$
wave is very close to 1 which means that the decay amplitudes calculated in our model 
are adequate.

Our value of $\chi$ parametrizes a large range of $K^+ K^-$ effective mass up to 1.8 GeV
and not just the region of $f_0(980)$. Therefore it cannot be directly compared with the 
value given in Ref.~\cite{Furman}. In addition, the estimate of $\chi$ given in Eq. (18) of ~\cite{Furman}
involves the coupling constant of $f_0(980)$ to $\pi \pi$ while here we have coupling to
$ K K$. Using $g_{f_0 \bar K K}/g_{f_0\pi \pi}=4.2$ from Ref.~\cite{BES}, a very rough estimate
 similar to that given in Ref.~\cite{Furman} leads to $\chi \approx 5.6$ GeV$^{-1}$.The $\kappa$
parameter was not used in Ref.~\cite{Furman} where only the on-shell contributions to the form 
factors were taken into account. The value of $\kappa=3.51$ GeV$^{-1}$ is reasonably 
larger than the typical $K K$ mass considered.
 We have also done an analogous fit to the data using the three $P$-wave form factors 
based on the parameterization of Ref.~\cite{Cheng} obtaining similar values of parameters as those 
written above, however with a higher $\chi^2$ value of 354.

 Fig. 1 shows the moduli of scalar
form factors $\Gamma^{n^*}_2(s_{23})$ and $\Gamma^{s^*}_2(s_{23})$ which determine the functional
dependence of the $S$-wave amplitudes on the $K^+K^-$ effective mass. There are two prominent maxima
of both form factors,
one related to the $f_0(980)$ resonance and the second one forming cusps due to the opening of the
third channel at 1400 MeV (in the present model responsible effectively for the production of four pions).
Presence of $f_0(980)$ leads to the threshold enhancement of the $S$-wave amplitude. This effect can be
directly studied in high statistics experiment with a very good effective $K^+K^-$ mass resolution of 
about 1 MeV and should be seen only a few MeV above the threshold.
\begin{figure}[t!]
\includegraphics[width=0.45\textwidth]{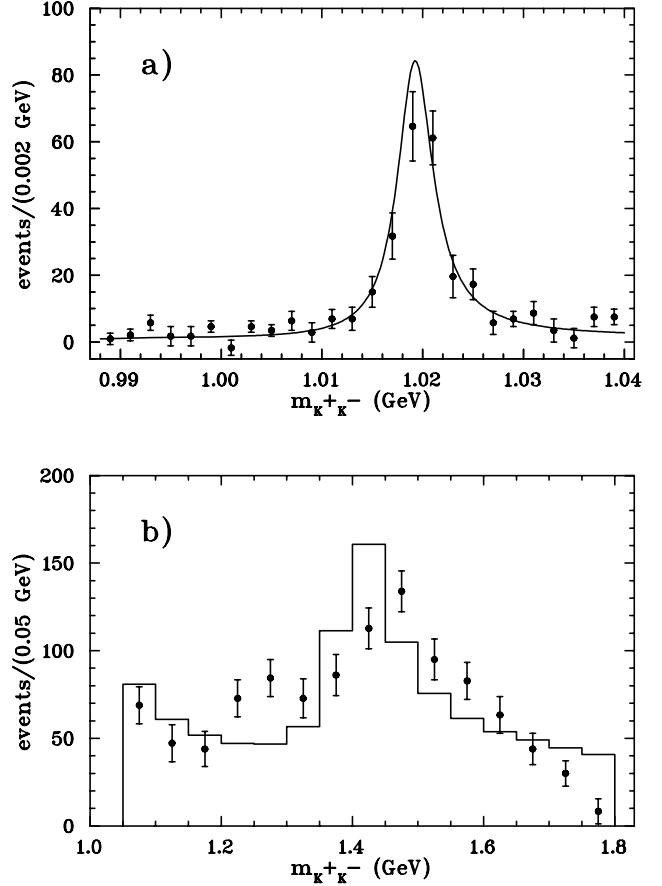}
\vspace*{6pt}
\caption{The $K^+K^-$ effective mass distributions from the fit to BaBar experimental
data \cite{BaBar} in the $\phi(1020)$ range (a) and between 1.05 GeV and 1.8 GeV (b). 
Theoretical results are shown as solid line in (a) and as histogram in (b).}
\end{figure}

In Fig.~2
the $K^+K^-$ effective mass distributions are shown for two mass ranges and for the data from
the BaBar Collaboration. At low $m_{K^+K^-}$ the spectrum is influenced by the $P$-wave amplitude and 
dominated by the $\phi(1020)$ resonance. Above 1.05 GeV the $S$-wave amplitude is much more important 
than the $P$-wave one. According to our analysis which uses the approach of Refs ~\cite{KLL,EPJC}, 
the experimental
 maximum near 1.5 GeV
can be attributed to the $f_0(1400-1460)$ found therein in solution $A$. We recall that in Ref. ~\cite{EPJC}
the coupling constant of the $f_0(1400)$ decay to $\bar K K$ is much smaller than the corresponding coupling
 to $\pi \pi$. Let us notice that
the model distribution depends on the sharp $4\pi$ threshold located at $2 \cdot m_3=1.4$ GeV which 
in reality should be smoothed out by the four-body pion interactions not taken into account in this 
quasi-two-body approximation. 
\begin{figure}[t!]
\includegraphics[width=0.45\textwidth]{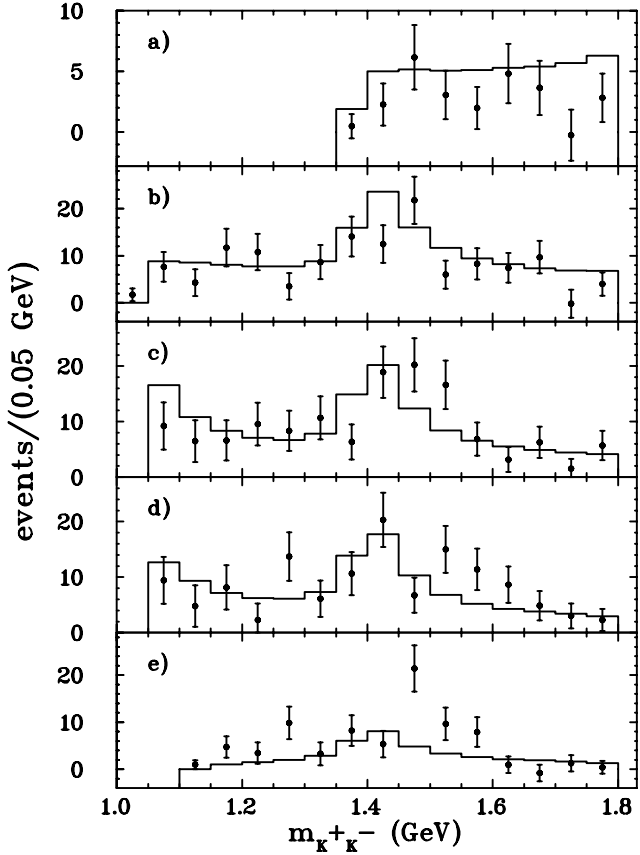}
\vspace*{6pt}

\caption{The $K^+K^-$ effective mass distributions from the fit to Belle experimental
data \cite{Belle} (a) for $m_{12}^2 < 5 $ GeV$^2$, (b) for $5$ GeV$^2 <m_{12}^2 < 10 $ GeV$^2$,
(c) for $10$ GeV$^2 <m_{12}^2 < 15 $ GeV$^2$, (d) for $15$ GeV$^2 <m_{12}^2 < 20 $ GeV$^2$ and
(e) for $20$ GeV$^2  <m_{12}^2$. Theoretical results are shown as histograms.}
\end{figure}
We have also studied the Belle~\cite{Belle} $K^+K^-$ effective mass spectra and found that
the quality of their description is similar to that shown in Fig. 2 for the BaBar data.
Fig.~3 shows a more detailed comparison of the $m_{K^+K^-}$ theoretical distributions with the Belle data 
\cite{Belle}, with events grouped
 in five ranges of $m_{12}$ which is the other combination of the $K^+K^-$ effective masses.
One observes an overall general agreement of theoretical histograms with experiment, with some
surplus of experimental events in Fig.~3e for the case of the highest slice of $m_{12}$ (larger than 20 
GeV$^2$)  where our model is not fully applicable due to the proximity of the Dalitz plot edge.

\begin{figure}[t!]
\includegraphics[width=0.4\textwidth]{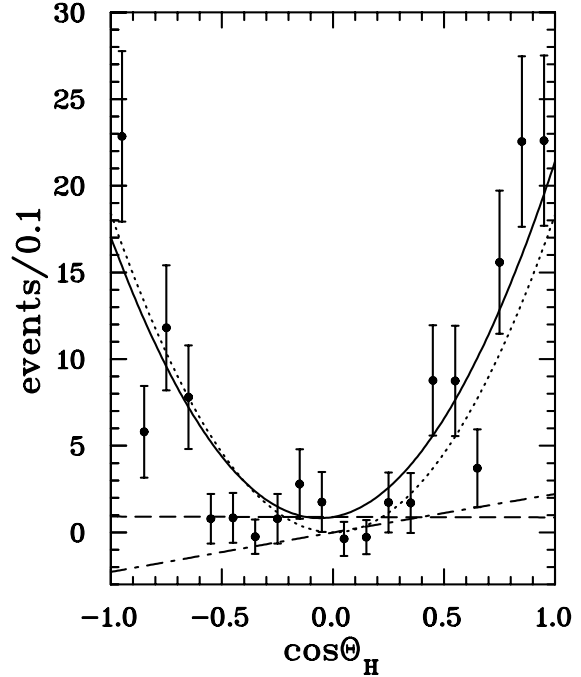}
\vspace*{6pt}

\caption{ 
Helicity angle distribution for the Belle data  \cite{Belle} in the $K^+K^-$ effective mass up
 to 1.05 GeV. The dashed line represents the $S$-wave contribution of our model, the dotted line 
- that of the $P$-wave, the dot-dashed - that of the interference term and the
solid line corresponds to the sum of these contributions.}
\end{figure}
Finally, in Fig.~4 we present the helicity angle distribution in the $K^+K^-$ mass range dominated by the
$\phi(1020)$ resonance. Without the $S$-wave component of the decay amplitude the distribution should be
symmetric with respect to $\ cos~\Theta_H =0$. However, we observe an interference effect which distorts
the distribution. This is a direct evidence of a non-zero part of the $S$-wave present even under the huge 
peak of the $\phi(1020)$ resonance. A theoretical integration of the $S$-wave contribution to the spectrum
in the $m_{K^+K^-}$ range from threshold till 1.05 GeV leads to about 12\% relative branching fraction.
It corresponds to the average branching fraction of $4.83\cdot 10^{-7}$ which is in agreement with the
 experimental upper bound of $2.9\cdot 10^{-6}$ found in Ref.~\cite{Belle}.
This agrees also with the BaBar estimate $(9 \pm 6) \%$ of the $S$-wave fraction in the region
of masses between 1.013 and 1.027 GeV \cite{BaBar}.
In the range of the $K^+ K^-$ effective mass from 1200 to 1800 MeV, which might be relevant for the 
$X_0(1550)$ discussed in Ref.~\cite{BaBar}, the $CP$ averaged branching fraction corresponding
 to the $S-$wave is equal to $4.42\cdot 10^{-6}$ which is larger than the total contribution of the $\Phi(1020)$
resonance.

We have also studied the $CP$ violation effects comparing the magnitudes of the decay amplitudes
of the $B^-$ and $B^+$ decays. While the moduli of the $P$-wave amplitudes for these charge conjugated decays
are rather similar, the $S$-wave amplitudes behave differently indicating an important $CP$
violation effect which depends on the $m_{K^+K^-}$ range. For the $S$-wave parameters written above, starting from 
the $K^+K^-$ threshold up to about 1.4 GeV, the modulus of the $B^+$ $S$-wave amplitude is larger than
the corresponding modulus of the $B^-$ amplitude. Then, above 1.4 GeV, the $B^-$ moduli become larger than
the $B^+$ ones. Defining the $CP$ asymmetry as
\begin{equation}
\label{ACP}
A_{CP}(m_{23})=\Big (\frac{dBr^-}{dm_{23}}-\frac{dBr^+}{dm_{23}}\Big )/
\Big (\frac{dBr^-}{dm_{23}}+\frac{dBr^+}{dm_{23}}\Big ),
\end{equation}
one gets very large asymmetries if one takes into account solely the contribution of the $S$-wave.
For example, $A_{CP}^S(1$ GeV$)=-0.51$, $A_{CP}^S(1.020$ GeV$)=-0.54$, 
$A_{CP}^S(1.25 $ GeV$)=-0.95$, and $A_{CP}^S(1.50 $ GeV$)=+0.59$ (here the superscript $S$ stands for the
$S$- wave asymmetry). When the $P$-wave is included then
the $CP$ asymmetry is reduced to:
$A_{CP}(1$ GeV$) = -0.25$, $A_{CP}(1.020$ GeV$)= +0.029$, $A_{CP}(1.25 $ GeV$)=-0.85$ and
$A_{CP}(1.50 $ GeV~$) = +0.495$. Let us note a particularly small asymmetry in the range of the
$\phi$ resonance, where the $P$-wave amplitude dominates, and an inversion of the $A_{CP}$ sign above 
1.4 GeV. Due to cancellations
between the ranges of the negative and positive asymmetries the resulting $CP$ asymmetry averaged
over the $m_{K^+K^-}$ range from threshold up to 1.8 GeV is rather small, equal to -0.05. 
The averaged
branching fraction for the same mass range equals to $9.6 \cdot 10^{-6}$. It is worthwhile to add
 that 
the $S$-wave gives to it the dominant contribution of $5.8 \cdot 10^{-6}$. 

\section{\bf Conclusions}
We have studied final state interactions between kaons in the $B^{\pm} \rightarrow K^+ K^- K^{\pm}$ 
decays. An overall general agreement with the Belle and BaBar data has been obtained. Our formalism
is based on the QCD factorization supplemented with the inclusion of the long distance $K^+K^-$ 
interactions. The latter are taken into account through the 
functional dependence of the scalar and vector form factors on the effective $K^+K^-$ masses.
A unitary model is constructed for the scalar non-strange and strange form factors in which three
scalar resonances $f_0(600)$, $f_0(980)$ and  $f_0(1400-1460)$ are naturally incorporated.  
 The scalar resonance $f_0(980)$ leads to the threshold enhancement of the $S$-wave $K^+K^-$ amplitude.
The $K^+K^-$ structure seen near 1.5 GeV can be attributed to the third scalar resonance. 
A potentially large $CP$ asymmetry is obtained in 
 the mass spectrum dominated by the $S$-wave. It originates from violent phase variations of the two kaon
 scalar form factors which affect 
the $K^+K^-$ effective mass dependence of the $S-$wave decay amplitudes. In general one can best study 
this effect away from the $\phi(1020)$ peak. 
We have shown, however, that even under the $\phi$ maximum one observes nonnegligible helicity angle asymmetry.
This effect originates from the interference between the $S$- and $P$- waves. 

Our approach presented here for the $B^{\pm} \rightarrow K^+ K^- K^{\pm}$ decays can be extended to study
the $B^0 \rightarrow K^+ K^- K^0_S$  reactions for which results of the time-dependent Dalitz analyses
have been recently published by the Babar~\cite{BaBar0} and
 Belle~\cite{Belle0} Collaborations .
For further studies of the charged $B$ decays new experimental data with better statistics are needed.  
Such data already
exist! For example, the Belle Collaboration has now five times larger data sample than 
that used in their publication \cite{Belle} analysed by us here. Future results from LHCb and 
from super-B factories would also be very useful.

\section*{Acknowledgments}
This work has been supported in part by the Polish Ministry of Science and
Higher Education (grant No. N N202 248135).

\end{document}